\documentclass[pra,aps,twocolumn,longbibliography, nofootinbib, showpacs]{revtex4-2}

\usepackage[colorlinks=true, citecolor=blue, urlcolor=blue, linkcolor = blue ]{hyperref}
\usepackage{graphicx,bm,amsmath,hyperref,xcolor,braket,plain,color,amsthm,amsfonts,float,tabularx,graphicx,bbm, mathtools,esvect,wrapfig,verbatim,enumitem,dcolumn,amssymb,appendix, physics,fmtcount,booktabs,csquotes}

\usepackage[normalem]{ulem}

\usepackage{tabularx}
\usepackage{latexsym}
\usepackage{graphics,epstopdf}
\usepackage{fmtcount}
\usepackage{booktabs}
\usepackage{csquotes}
\usepackage{epsfig}

\usepackage[mathscr]{euscript}

\newcommand{\Fmu}[0]{\mathcal{F}_{\mu\mu}}
\newcommand{\FDD}[0]{\mathcal{F}_{\Delta\Delta}}
\newcommand{\FmD}[0]{\mathcal{F}_{\mu\Delta}}
\newcommand{\Ek}[0]{\mathcal{E}_{k}}
\DeclareMathAlphabet{\mathpzc}{OT1}{pzc}{m}{it}

\begin{document}
\title{Multicritical quantum sensors driven by symmetry-breaking}


\author{Sayan Mondal, Ayan Sahoo, Ujjwal Sen and Debraj Rakshit}
\affiliation{Harish-Chandra Research Institute, A CI of Homi Bhabha National Institute,  Chhatnag Road, Jhunsi, Prayagraj 211 019, India}

\begin{abstract}
Quantum criticality has been demonstrated as a useful quantum resource for parameter estimation. This includes second-order, topological and localization transitions. In all these works reported so far, gap-to-gapless transition at criticality has been identified as {a crucial resource} for achieving the quantum-enhanced sensing, although there are several important concepts associated with criticality, such as long-range correlation, symmetry breaking. {In this work, we show that symmetry-breaking alone can drive a quantum-enhanced sensing, even without any gap-to-gapless transition. We analytically demonstrate that the estimation of the superconducting pairing amplitude in the one-dimensional Kitaev model achieves Heisenberg scaling when the system is prepared near a multicritical point and is varied along a gapless critical line, implying symmetry breaking as a standalone metrological resource. Extending our analysis in the realm of simultaneous multiparameter estimation of both the pairing term and the chemical potential, we show that it is possible to obtain $L^6$ scaling in a narrow parameter range, but with definite observable consequence, where the quantum advantage is assisted by gap-to-gapless transition as well. Our work thus identifies a new resource for criticality-enhanced quantum sensing, and also suggests multicritical systems as useful platform for multiparameter sensing.}  

\end{abstract} 

\maketitle

\section{Introduction}
The quantum Cram{\'e}r-Rao bound  provides the ultimate limit of precision within which an unknown parameter can be estimated in the quantum systems \cite{Braunstein1994,Degen2017,Cramer1946,1969JSP.....1..231H}. The uncertainty $\epsilon_q$ in parameter estimation is related to a very special quantity, quantum Fisher information (QFI). The bound suggests that  $\epsilon_q \ge (M F_Q)^{-1}$ for the quantum systems, where $F_Q$ is the QFI and $M$ is the number of repetition of the sensing protocol. Quantum systems might provide a true quantum edge over classical systems when it comes to accurately measuring an unknown variable. This is demonstrated by how the uncertainty scales with the size of the probe. Given $\epsilon_q \propto L^{-\nu}$, where $L$ is the probe size and $\nu$ is the associate scaling exponent,  the classical probes can at-best scale linearly with system size, i.e., $\nu=1$, which corresponds to the standard quantum limit (SQL). However, it is possible to exploit quantum phenomena, such as quantum correlation or cooperative phenomena, for achieving quantum-enhanced sensitivity, where $\nu > 1$. The so-called Heisenberg limit (HL) corresponds to $\nu = 2$ \cite{giovannetti2004quantum}. 
Subsequently, quantum sensing \cite{giovannetti2011advances,Degen2017,PhysRevLett.132.100803}  has taken the centre stage of quantum technology with several applications
, and has been investigated from different points of views \cite{giovannetti2004quantum,PhysRevLett.96.010401,PhysRevE.101.052107,manshouri2024quantum,Chiranjib23,bhattacharyya2024even,yousefjani2024discrete}.

The cooperative quantum phenomena, such as quantum phase transitions in quantum many-body (QMB) systems, has turned out to be an extremely useful quantum resource for engineering new types of quantum sensors. 
In literature, quantum criticality has been used to undertake various metrological tasks \cite{Frerot2018, Garbe2020, Hotter2024, Agarwal2025_QuantumSensingReview}.
Quantum criticality, in its various forms, have been utilized to attain HL limit, or for even realizing beyond HL limits. This includes symmetry-breaking second-order quantum phase transitions \cite{Zanardi2006,PhysRevB.77.245109,Zanardi2008,PhysRevA.78.042106,Gu2010,gammelmark2011phase,skotiniotis2015quantum,PhysRevX.8.021022,PhysRevA.99.042117,PhysRevLett.126.010502,huggins2021efficient,mirkhalaf2021criticality,di2023critical,PhysRevLett.130.240803,Mishra2021, Montenegro2021, Montenegro2022, Monika2023, Singh2024,Dooley2023,Montenegro2023,Gu2010, Pezze2019, Ilias2022, Yang2023}, as well as symmetry-protected quantum phase transitions, such as localization transition \cite{He2023,Sahoo2024, sahoo2024stark, Debnath2025_TiltInducedLocalizationBEC} and topological phase transition (TPT) \cite{PhysRevLett.120.250501,PhysRevLett.120.250501,PhysRevB.102.224401,PhysRevB.100.184417,PhysRevResearch.4.023144,PhysRevResearch.4.013133,Sarkar2022}, and non-Hermitian topological systems \cite{PhysRevLett.112.203901,PhysRevResearch.2.013058,PhysRevLett.125.180403,PhysRevResearch.4.013113}.
{It has been demonstrated in literature that when considering other resources like preparation time, the apparent super-HL scaling can diminish to HL \cite{PhysRevX.8.021022, PhysRevLett.132.100803}. However, recently it has been shown that even after considering preparation time, super-HL scaling can be preserved in certain systems \cite{He2023}. The effectiveness of super-HL scaling quantum sensing based on QFI and quantum Cram\'er-Rao bound has been studied extensively in the literature and it has been found that sub-HL strategies are ineffective \cite{Giovannetti2012a, Giovannetti2012b}.}

There are two different key motivations for probing different kinds of quantum criticality - first, it is important to understand the basic ingredients leading to quantum-enhanced sensitivity, e.g. the role of gap closing or symmetry-breaking; secondly, there is a constant quest for identifying experimentally realizable QMB systems with super-Heisenberg scaling property characterized by large scaling exponent.  It has been shown that the gap closing at criticality leads to a scaling of $F_Q \sim L^{(2/\nu)}$,  where $L$ is the system size, and  the critical exponent $\nu$ characterizing the divergence of the correlation length  (localization length) near the criticality in case of the second-order quantum phase transition (localization transition).  Although there are many key aspects associated with quantum criticality, such as symmetry breaking, gap closing, long-range correlation, only the gap closing has been identified as the crucial ingredient for obtaining quantum-enhanced sensitivity, implying $\nu < 2$. { It remains unclear whether other critical properties, such as symmetry breaking, can independently provide quantum advantage.}

In this work we show that gap-to-gapless transition is  not always required for quantum-enhanced sensing. {We show that symmetry breaking alone, when the system is prepared near a multicritical point, can be sufficient to obtain Heisenberg-limited scaling in single-parameter estimation.} To the best of our knowledge, such symmetry-breaking-assisted quantum advantage for quantum sensing has not been reported in literature previously. { We establish this result while performing single-parameter estimation of the superconducting pairing gap in the one-dimensional Kitaev model. We then extend the analysis to simultaneous multiparameter estimation of the pairing term and the onsite potential, where multicriticality allows super-Heisenberg scaling. The individual contributions of symmetry breaking and gap closing, however, can no longer be separated, as both effects coexist.}

{The rest of the paper is organized as follows. In Sec.~\ref{Sec2}, we provide a brief overview of the Kitaev model on one-dimensional lattice and briefly discuss the theory of multi-parameter sensing.  {In Sec.~\ref{sec:single} and \ref{sec:onsite}, we consider single-parameter sensing. There in Sec.~\ref{sec:single} we demonstrate how symmetry-breaking can be exploited as a resource for quantum sensing of the pairing amplitude and establish one of the central results of this work. In Sec.~\ref{sec:multi}, the QFI matrix for multiparameter sensing is presented. We demonstrate that  multicritical point can be exploited for quantum enhanced sensing of multiple parameters, representing another main result of this work.} In Sec.~\ref{Sec4}, we study the parameter regime in which the proposed sensor provides quantum advantage. We also examine how the sensor works, when state preparation time is taken into consideration. Finally, we present our concluding remarks and discussions in Sec.~\ref{Conc}.}

\section{Preliminaries}
\label{Sec2}
\subsection{Model}
\begin{figure}
    \centering    \includegraphics[scale = 0.45]{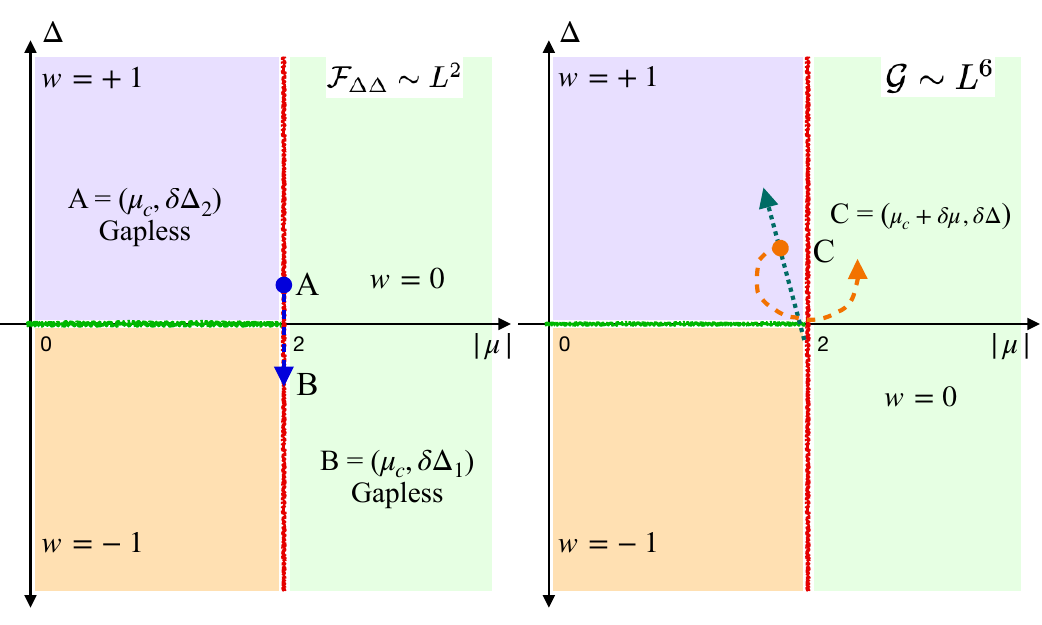}    \caption{\textit{\bf{Schematic for quantum-enhanced sensing assisted by symmetry-breaking}}: The phase diagram shows different topological phases in the $\mu$-$\Delta$ plane.  The red and the green lines are gapless. In the left panel, the protocol of sensing $\Delta$ is presented. $\FDD$ has an quantum-enhanced scaling when the system is prepared at the critical line of $\mu=\mu_c$ (near the multicritical point, at A), and driven to, say B. We note that this is enhancement of scaling is not because of gap-to-gapless transition, but rather due to breaking of $U(1)$ symmetry. In the multi-parameter regime, one can envision a case where system is prepared at C, and both the parameters are maneuvered along the directed arrows. 
    {\color{black}Since, we are in a two-dimensional plane, the point C can be approached from multiple directions.} 
    This leads to a quantum-enhanced scaling for the multi-parameter precision estimator.
    This involves, both, a gap-to-gapless transition and the symmetry-breaking. }
    \label{fig:schematic}
\end{figure}

We consider the one-dimensional Kitaev lattice model~\cite{Kitaev_2001,Mbeng2024}, $\hat{H} = \hat{H}_1 + \hat{H}_{\Delta}$, where
\begin{align}
    \label{eq:kitaev}
    \hat{H}_1 &= -\sum_{j = 1}^{L} \left(\hat{c}_j^\dagger \hat{c}_{j+1} + h.c.\right)-\mu \sum_{j = 1}^L \left( \hat{c}_j^\dagger \hat{c}_{j} - \frac{1}{2}\right), \nonumber \\
    \hat{H}_{\Delta} &= \frac{\Delta}{2} \sum_{j = 1}^{L} \left( \hat{c}_j^\dagger \hat{c}^\dagger_{j+1} + h.c.\right).
\end{align}
Here $\hat{c}^{\dagger}_j$ ($\hat{c}_j$) is the creation  (annihilation) operator at site $i$ of a spinless fermion, $L$ is the number of lattice sites, $\mu$ is an on-site potential and $\Delta$ is the superconducting $p$-wave pairing amplitude. This model has a rich topological phase diagram~\cite{Leijnse2012, Elliot2015,Maity2019,Pezze2017}, which can be characterized via a topological invariant, the winding number, $w$. Two critical lines $\mu= \pm 2$ and $\Delta = 0$ set apart three topologically distinct phases: a trivial phase for $|\mu| > 2$ with $w=0$, one topologically non-trivial phase for $|\mu| < 2$ with $w=1$ for $\Delta > 0$, and another topologically non-trivial phase for $|\mu| < 2$ with $w=-1$ for $\Delta < 0$. 
 A schematic representation of these phases and the associated critical lines is shown in Fig.~\ref{fig:schematic}. 
The one-dimensional Kitaev model hosts Majorana Zero Modes \cite{Hasan2010, Asboth2016} in the topological phases at both of its ends. The intersections of two critical lines mark two multicritical points in the system. The multicritical point connects two topologically non-trivial phases with a topologically trivial phase. Topological properties of a $p$-wave spin-less superconductor  have been studied from different points of views in recent years \cite{PhysRevB.83.155429,Sau2012,PhysRevB.84.214528,degottardi2011topological,DeGottardi2013,Thakurathi2013,Maity2019,Fraxanet2021}. There has been several proposals and experimental efforts for physical realization of the Kitaev $p$-wave model \cite{Fu2008,Sau2012,mourik2012signatures,Deng2012,Das2012, Fulga_2013,  Nadj-Perge2014, Xu2016, Xing2018, Dvir2023,Liu2022, Zhang2023}.

\emph{Many-body ground state}.-- The ground state of the Kitaev model under anti-periodic boundary conditions (ABC) is a many-particle state which is given by,
\begin{align}
    |\Psi_G\rangle = \prod_{k = 1}^{L/2} (u_k + v_k c_k^\dagger c_{-k}^\dagger) |0\rangle,
    \label{eq:phi_G}
\end{align}
where $ u_k = (\mathcal{E}_k + z_k)/E_k$ and  $v_k ={iy_k}/E_k$. Here 
$E_k={\sqrt{2 \mathcal{E}_k(\mathcal{E}_k + z_k)}}$ with $\mathcal{E}_k = 2\sqrt{(\cos k + \frac{\mu}{2})^2 + \frac{1}{4}\Delta^2 \sin^2k}$, $y_k = -\Delta \sin k$ and $z_k = -\mu - 2\cos k$. The ground state is a superposition of states with multiple particle numbers but all are from even parity sector. 

\subsection{Theory of single- and multi-parameter sensing}
\label{Sec3}
Let us consider a set of parameters, $\mathbf{x} = \{x_a\}_{a = 1}^n$, encoded in some density matrix $\rho(\mathbf{x})$. The precision limits of any parameter that is being sensed is given by a $n \times n$ matrix called the quantum Fisher Information matrix (QFIM). The elements of the QFIM represented by $\mathcal{F}$ is given by, $\mathcal{F}_{ab} := \frac{1}{2}\text{Tr}\left( \rho\{L_a,L_b\}\right)$,
where $\{\cdot,\cdot\}$ represents the anti-commutation and the operators $L_{a}$ and $L_{b}$ are the symmetric logarithmic derivatives (SLD) with respect to $x_a$ and $x_b$, respectively. This is given by the equation,  $\partial_i \rho = \frac{1}{2}\left( \rho L_i + L_i\rho\right)$,
where, $i = a,b$ and $\partial_i$ is the partial derivative with respect to $x_i$ viz. $\partial_i = \partial/\partial x_i$. 
In this work, we specifically work with the many-body ground state of the Kitaev model. Hence our state of interest is a pure state. For pure states, the elements of QFIM reduces to  $\mathcal{F}_{ab} = 4 \text{Re}\left( \langle \partial_a\psi|\partial_b\psi\rangle - \langle \partial_a\psi|\psi\rangle \langle\psi|\partial_b\psi\rangle \right)$.    
Here the state $|\psi\rangle = |\psi(\mathbf{x})\rangle$ encodes the parameters $\mathbf{x}$. It can be seen that the diagonal elements of QFIM - $\mathcal{F}_{aa}$ reduces to the QFI for parameter $x_a$. 

The parameters $\mu$, $\Delta$ of the hamiltonian in Eq.~\eqref{eq:kitaev}, are encoded in the ground state. The adiabatic changes in these parameters gets reflected in the ground state. By employing the methods of the QFIM, we get a $2\times2$ matrix $\mathcal{F}$, given by
\begin{align}
    \mathcal{F} =   \begin{pmatrix} 
                    \mathcal{F}_{\mu \mu} & \mathcal{F}_{\mu \Delta} \\
                    \mathcal{F}_{\Delta \mu} & \mathcal{F}_{\Delta \Delta}
                    \end{pmatrix}.
\label{eq:QFIM-app}
\end{align}
Recently, multi-parameter sensing has gained a lot of interest~\cite{DiFresco2022, Mihailescu2024, Yousefjani2024}. The precision limit of a multi-parameter estimation is quantified using the covariance matrix, $\text{Cov}(\mathbf{x})$. This is given by 
\begin{align}
\text{Cov}_{ij}(\mathbf{x}) := \langle x_i x_j \rangle - \langle x_i \rangle \langle x_j\rangle.    
\end{align}
The QFIM, $\mathcal{F}$ is related to the covariance matrix $\text{{Cov}}(\mu, \Delta)$ and obeys following Cram\'er-Rao inequality:  
\begin{align}
\text{Cov}(\mu, \Delta) \geq \frac{1}{m}\mathcal{F}^{-1}(\mu, \Delta).
\label{eq:CRB}
\end{align}
Taking trace on both sides of the matrix inequality one obtains the scalar inequality,
\begin{align}
\label{eq:multi-presc-app}
    \delta \mu^2 + \delta \Delta^2 \geq \frac{1}{m}\frac{\mathcal{F}_{\mu \mu} + \mathcal{F}_{\Delta \Delta}}{\mathcal{F}_{\mu \mu}\mathcal{F}_{\Delta \Delta} - \mathcal{F}_{\Delta \mu}\mathcal{F}_{\mu \Delta}} = \frac{1}{m} \mathcal{G}^{-1},
\end{align}
where we define, 
\begin{align}
\mathcal{G} := (\text{Tr}[\mathcal{F}^{-1}])^{-1}=\frac{\mathcal{F}_{\mu \mu}\mathcal{F}_{\Delta \Delta} - \mathcal{F}_{\Delta \mu}\mathcal{F}_{\mu \Delta}}{\mathcal{F}_{\mu \mu} + \mathcal{F}_{\Delta \Delta}}.
\label{def_G-app}
\end{align}
The quantity $\mathcal{G}^{-1}$ is the total uncertainty for the equally weighted multi-parameter estimation and $m$ is the number of repetition of the sensing protocol. For single-parameter estimation, where both parameters are sensed individually and independently, this precision bound reduces to,
\begin{align}
\label{eq:single-presc}
    \delta \mu^2 + \delta \Delta^2 \geq \frac{1}{m}\left(\frac{1}{\mathcal{F}_{\mu \mu}} + \frac{1}{\mathcal{F}_{\Delta \Delta}}\right).
\end{align}
Note that the off-diagonal terms are symmetric under interchange of $\mu$ and $\Delta$, i.e., $\mathcal{F}_{\Delta \mu} = \mathcal{F}_{\mu \Delta}$. This implies that right-hand side quantity in Eq.~\eqref{eq:multi-presc-app} is smaller than that in Eq.~\eqref{eq:single-presc}, and hence thereby making the Eq.~\eqref{eq:multi-presc-app} a tighter bound than the Eq.~\eqref{eq:single-presc}. Thus, we observe that multi-parameter estimation admits same uncertainty as separate single-parameter estimations, if and only if the parameters under study are uncorrelated.

For the single-parameter estimation case, i.e., when all the parameters are assummed to be known except one, say $\theta$, the estimation procedure reduces to computing the diagonal entry of the QFIM corresponding to the parameter of interest. This non-zero element alone determine the quantum Crame{e'}r-Rao bound: $\Delta \theta \ge 1/ \sqrt{F_{\theta\theta}}$ (see also Appendix \ref{app:A}). The off-diagonal and remaining entries are irrelevent in this context. Thus, for single-parameter estimation, the relevant diagonal element of the QFIM fully characterizes the ultimate metrological precision.

\begin{figure*}[t]
    \centering
    \includegraphics[scale = 0.4]{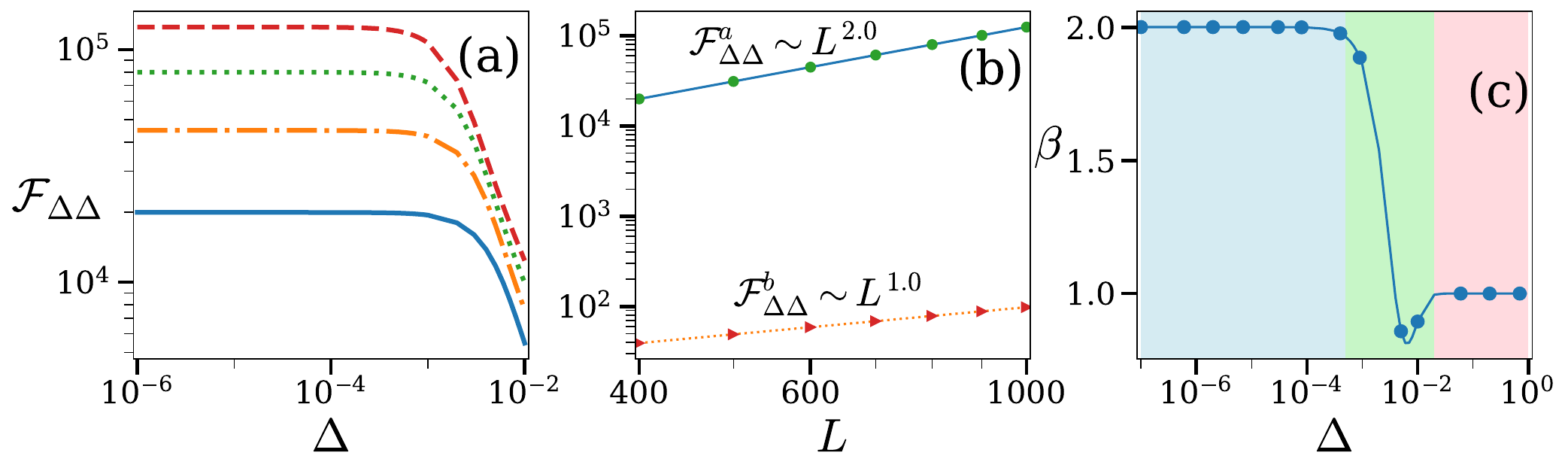}    \caption{\textit{\textbf{Variation of $\FDD$ near the multicritical point}}: (a) $\FDD$ for system sizes $L$ = 400 (solid blue line), 600 (dot-dashed orange line), 800 (dotted green line), 1000 (dashed red line) are presented with respect to $\Delta$ keeping a fixed $\mu = 2$. This is a representative case, we observe similar nature for various other system-sizes as well. The finite-size scaling of $\FDD$ is presented in (b). Setting $\mu = 2$, $\FDD$ at $\Delta = 10^{-7}$ (circular markers) scale as $\FDD^a \sim L^2$ while at $\Delta = 0.7$ (square markers) they scale as $\FDD^b \sim L$. The straight lines are the best fit. 
    (c) We present the scaling exponents $\beta$ of $\FDD$ against the parameter $\Delta$. We identify three regions, (i) low $\Delta$ regime: $\Delta < 5\times10^{-4}$ is represented by light blue color where $\FDD \sim L^2$; (ii) intermediate $\Delta$ regime : $5\times10^{-4}\leq\Delta\leq0.02$ represented by light green color where $\beta$ transitions; (iii) large $\Delta$ regime: $\Delta > 0.02$ is represented by light pink color where $\FDD \sim L$. 
    }
    \label{fig:FDD-G}
\end{figure*}
\section{Estimation of pairing amplitude}
\label{sec:single}
{ Although we later venture into the multiparameter estimation regime, it's necessary to isolate the case of single-parameter estimation of the pairing amplitude, $\Delta$, in order to establish symmetry-breaking as a standalone metrological resource. Unlike the second-order quantum phase transitions, the localization-delocalization transitions do not belong to the Landau paradigm of quantum phase transition, and do not involve symmetry-breaking at the critical point. Hence, gap-closing is singled out as the resource for quantum enhanced sensing of the critical sensors based on such transitions \cite{He2023,Sahoo2024, sahoo2024stark, Debnath2025_TiltInducedLocalizationBEC}. Similar is the situation in critical sensor based on the topological transition,  proposed in \cite{Sarkar2022}. In our model, the single-parameter setting helps in providing the fundamental understanding that symmetry-breaking itself can be an independent source of enhanced sensitivity, whereas in the multiparameter case, which we address later, its role becomes inseparably linked to critical gap closing.}

{ Let us first consider single-parameter estimation of $\Delta$, for which the onsite potential $\mu$ is considered to be a known control parameter.  This requires computation of the diagonal element of the QFIM (see Eq.~3) correponding to the unknown parameter $\Delta$, i.e., $\mathcal{F}_{\Delta\Delta}$. We evaluate $\mathcal{F}_{\Delta\Delta}$, when the ground state $|\Phi_G\rangle$ (Eq.~\eqref{eq:phi_G}) of the Kitaev model is used as a probe. Using the fidelity-based definition of $\mathcal{F}_{ab}$ discussed above, it turns out to be  
\begin{align}
\FDD &= \sum_k \frac{z_k^2 \sin^2k }{\mathcal{E}_k^4}, 
\label{eq:QFIM-ele-D}
\end{align}
where $z_k=-2-\cos(k)$. We set the control parameter at $|\mu|=2$, which corresponds to the gapless critical line. The parameter variation of $\Delta$ is then bound to happen along this critical line. Preparing the state near the multicrical point, the leading order behavior of $\FDD$ in the limit $\Delta \to 0$ turns out to be, 
\begin{equation}
    \mathcal{F}_{\Delta\Delta} \sim \frac{L^2}{8},
\end{equation}
implying a Heisenberg scaling  of the quantum Fisher information corresponding to the unknown parameter at multicriticality. The detailed derivation is presented in the Appendix~\ref{app:A}.} 

\subsection{Symmetry-breaking as a resource}
We present a set of exact results displaying the behavior and scaling of $\FDD$ with respect to the $\Delta$ in Fig.~\ref{fig:FDD-G}(a-c).  In Fig.~\ref{fig:FDD-G}(a) we present how $\FDD$ varies with the parameter $\Delta$ when $\mu = 2$ for various system-sizes. We observe that $\FDD$ saturates to a particular maximum value as $\Delta\rightarrow0$. The quantity $\FDD$ at $\mu = 2$ becomes independent of $\Delta$ for small finite values of $\Delta$, say up-to $\Delta^*$. As the value of $\Delta$ increases, $\FDD$ decreases gradually with increase in $\Delta$. It can be expected that $\Delta^* \to 0$ in the thermodynamic limit. In Fig.~\ref{fig:FDD-G}(b) the finite-size scaling of $\FDD$ is presented at $\mu = 2$ for two representative cases, $\Delta = 10^{-7}$ with Heisenberg scaling $\beta = 2$, and $\Delta = 0.7$ with SQL scaling $\beta = 1$, with $\beta$ being the scaling exponent. In Fig.~\ref{fig:FDD-G}(c), this transition of the system-size scaling from $L^2$ to $L$ of $\FDD$ with change in $\Delta$ is presented. 

{ The Heisenberg scaling in the estimation of weak pairing amplitude occurs despite the fact that the parameter variation of $\Delta$ occurs along the gapless line $|\mu|=2$, and thus eliminates gap-to-gapless transition as a probable resource.}  In absence of the pairing term, i.e., for $\Delta=0$, the  system has the global $U(1)$ symmetry as the Hamiltonian is invariant under the transformation $\hat{c}_i \to e^{i \phi} \hat{c}_i$, where $\phi$ is a site-independent phase. This global symmetry implies the conservation of the total fermion number. The corresponding system is gapless for parameter regime $|\mu| \le 2$ and $\Delta = 0$. Introduction of $\hat{H}_\Delta$ breaks the $U(1)$ symmetry, and a gap opening occurs. The phase diagram is gapped everywhere in the $\mu-\Delta$ plane except at the critical lines at $|\mu|=2$ and $\Delta=0$. The basic idea of the symmetry-breaking quantum enhancement of precision is the following: for parameter estimation of $\Delta$, assuming $\mu$ as a known parameter that can be precisely tuned, the many-body ground state of the system is prepared near the multicritical point ($\mu = 2$ and $\Delta \to 0$). Because the phase boundary is given by $\mu = 2$, any small change in $\Delta$ leaves the system on the associated gapless critical line. Consequently, the Fisher information $\FDD$ is evaluated along a gapless line, which isolates the gap-to-gapless transition as a resource and singles out instead the role of $U(1)$ symmetry-breaking. 

We take a closer look at this scenario. The ground state configuration of the system at the multicriticality corresponds to a fully-filled situation, implying a band-insulating configuration. The symmetry-breaking is physically associated with a band-insulating to a $p$-wave superconducting transition. For finite system-size, we observe an extended region, where the system effectively remains in the band-insulating regime and the effect of symmetry-breaking is not observed. This inference can be made by tracking the average particle number of the ground state $\langle N\rangle$. 
Further details on this is provided in Appendix \ref{avg-num-state}.  The situation is closely analogous to that reported in Ref.~\cite{He2023}, where the presence of a Stark field gives rise to an extended delocalized regime accompanied by super-Heisenberg scaling of the QFI. In contrast, for the single-parameter estimation of $\Delta$, the finite-size extended region exhibits Heisenberg scaling, although the underlying mechanisms in the two cases are entirely distinct.

\begin{figure}[t]
    \centering
    \includegraphics[scale = 0.37]{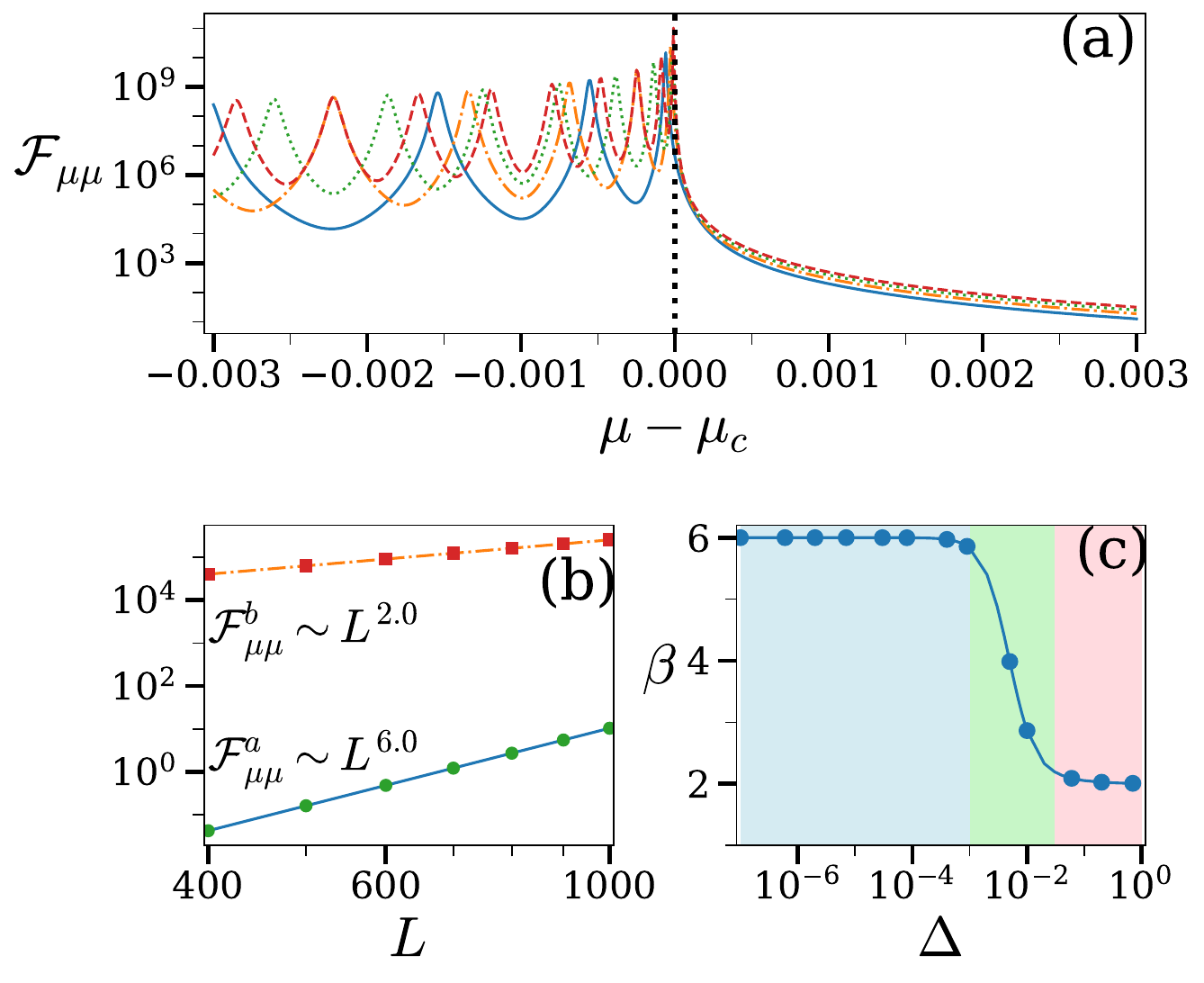}
    \caption{\textit{\textbf{The variation of $\Fmu$ around the multicriticality}} : (a) $\Fmu$ for system sizes $L$ = 400 (solid blue line), 600 (dot-dashed orange line), 800 (dotted green line), 1000 (dashed red line) are presented against $\mu$, where in the x-axis $\mu_c = 2$ has been subtracted from $\mu$. $\Delta = 0.001$ is set for all cases. This is a representative case, we observe a similar nature for various other system-sizes. (b) The finite-size scaling of $\Fmu$ is presented at $\mu = 2$ and two different $\Delta$. Considering $\Delta = 10^{-7}$, we observe $\Fmu^a \sim L^{6}$ (circular markers) and $\Delta = 0.7$, we observe $\Fmu^b \sim L^{2}$ (square markers). The straight lines are the best fit. (c) We present the scaling exponents $\beta$ of $\Fmu$ against the parameter $\Delta$. We identify three regions, (i) low $\Delta$ regime: $\Delta < 10^{-3}$ represented by light blue color where $\Fmu \sim L^6$; (ii) intermediate $\Delta$ regime: $10^{-3}\leq\Delta\leq10^{-2}$ is represented by light green color, where $\beta$ transitions from $6$ to $2$; (iii) large $\Delta$ regime: $\Delta > 10^{-2}$ is represented by light pink color, where $\Fmu \sim L^2$. 
    }
    \label{fig:Fmm}
\end{figure}

\begin{figure*}[t]
    \centering
    \includegraphics[scale = 0.45]{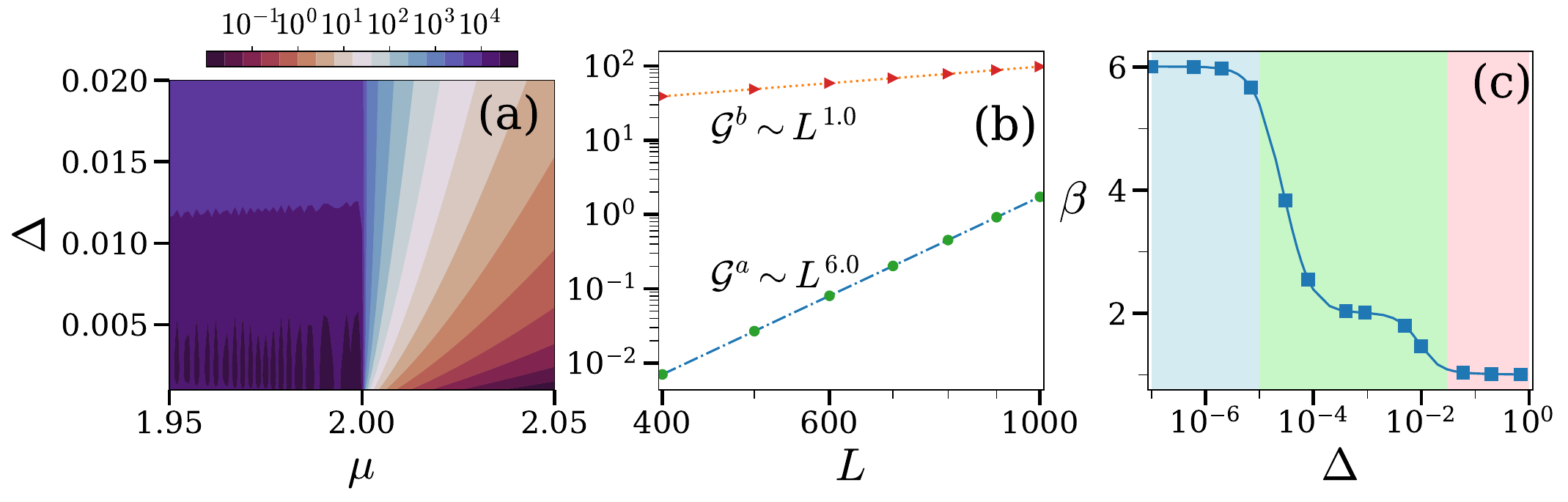}
    \caption{{\textit{\bf{Variation of $\mathcal{G}$ near multicriticality}}:  (a) The quantity $\mathcal{G}$ is presented with respect to the tuning parameters $\mu$ and $\Delta$ in the vicinity of the multicriticality. 
    The system size is $L = 1000$. 
    (b) The finite-size scaling of $\mathcal{G}$ is presented. Setting $\mu = 2$, $\mathcal{G}$ at $\Delta = 10^{-7}$ (circular markers) scale as $\mathcal{G}^a \sim L^6$, while at $\Delta = 0.7$ (square markers) we have $\mathcal{G}^b \sim L$. The straight lines are the best fit.
    (c) We present the scaling exponents $\beta$ of $\mathcal{G}$ against the parameter $\Delta$. We identify three regions, (i) low $\Delta$ regime: $\Delta < 10^{-5}$ is represented by light blue color where $\mathcal{G} \sim L^6$; (ii) intermediate $\Delta$ regime : $10^{-5}\leq\Delta\leq0.03$ represented by light green color where $\beta$ transitions for both $\FDD$ and $\mathcal{G}$; (iii) large $\Delta$ regime: $\Delta > 0.03$ is represented by light pink color where $\FDD, \mathcal{G} \sim L$.}}
    \label{fig:G}
\end{figure*}

\section{Estimation of onsite potential}
\label{sec:onsite}
{Before we proceed to the multiparameter estimation, let us also consider single-parameter estimation of $\mu$. This requires computation of the diagonal element of the QFIM (see Eq.~\eqref{eq:QFIM-app}) corresponding to the unknown parameter $\mu$, i.e., $\mathcal{F}_{\mu\mu}$. This situation, however, is different as the parameter $\Delta$ is considered to be the known control parameter.

 We obtain following expression for $\mathcal{F}_{\mu\mu}$, 
\begin{align}
\Fmu &= \sum_k \frac{\Delta^2 \sin^2 k}{\mathcal{E}_k^4}.
\end{align}
Around the multicritical point, the leading order behavior of $\Fmu$ is,
\begin{align}
    \Fmu \sim \frac{\Delta^2}{\pi^6}L^6.
    \label{eq:Fmm-1}
\end{align}
A complete derivation of the QFI, $\Fmu$, is given in Appendix~\ref{app:A}.  $\Fmu$ scales as $L^6$ near the multicritical point. It has a quadratic dependence on $\Delta$ and disappears at the multicritical point. In fact it vanishes along the entire critical line $\Delta=0$, which is in complete contrast to $\mathcal{F}_{\Delta\Delta}$ that achieves quantum enhanced sensitivity near the multicritical point.
It is necessary to set $\Delta$ to a  finite value in order to obtain finite $\Fmu$.}

Setting $\Delta$ at  finite non-zero values near the multicritical point and moving across the critical line ($\mu = 2$) we observe that maximum value is observed at $\mu = 2$. This is presented in Fig.~\ref{fig:Fmm}(a), for various system sizes, with $\Delta$ being set at $0.001$. In Fig.~\ref{fig:Fmm}(b), we present the finite-size scaling of $\Fmu\sim L^\beta$, at $\mu = 2$ and for two representative cases of $\Delta = 10^{-7}$ with scaling exponent $\beta = 6$ and $\Delta = 0.7$ with $\beta = 1$. This change in the scaling exponent happens gradually with respect to the change in the value of $\Delta$ and is presented in Fig.~\ref{fig:Fmm}(c). We observe that based on the value of $\beta$, the parameter range of $\Delta$ can be divided into three parts: small $\Delta$ regime with super-Heisenberg scaling ($\beta = 6$), intermediate $\Delta$ regime where a transition happens from super-HL to SQL ($\beta = 1$),  and large $\Delta$-regime where we observe SQL scaling.

At $\Delta = 0$, $\Fmu$ vanishes, but at small but at non-vanishing values it provides a $L^6$ scaling with system-size, implying that it is essential to have $\Delta \neq 0$, for enhanced quantum sensing. The Eq.~\eqref{eq:Fmm-1} is obtained by setting $\mu = 2$ and expanding it for $\Delta \to 0$ in the expression of $\Fmu$ in Eq.~\eqref{eq:QFIM-ele}. We observe that the  $\Fmu\sim L^6$ scaling is valid only up to certain system-size $L_c$, beyond which it scales as $\Fmu\sim L^2$. We observe that for a specific $\Delta$ value, $L_c \sim \Delta ^{-1}$, thus as $\Delta$ decreases in value, $L_c$ increases, making the super-HL scaling valid for larger system sizes. This is further discussed in detail in Appendix~\ref{app:A} and Fig.~\ref{fig:L-change}. 

{ In contrast to the $\Delta$-estimation scenario, quantum-enhanced sensing of $\mu$ necessitates a gap-to-gapless transition. It is also essential to stay within the symmetry-broken phase. Thus, the combined role of  gap-to-gapless transition and the symmetry-breaking act as resource in this case.  
}

\section{Multi-parameter Estimation}
\label{sec:multi}
{
Finally, we investigate the multi-parameter sensing of the one-dimensional Kitaev model near the multicritical point. It addresses simultaneous estimation of both the parameters $\mu$ and $\Delta$. As discussed in section~\ref{Sec3}, the quantity $\mathcal{G}^{-1}$ bounds the total uncertainty for the equally weighted multi-parameter estimation. We calculate the QFIM elements in Eq.~\eqref{eq:QFIM-app}, when the ground state $|\Phi_G\rangle$ of the Kitaev model is used as a probe. {The expressions for the diagonal elements have already been reported, and we are left with evalution of the off-digonal element, $\mathcal{F}_{\mu \Delta}$.}

Following expressions are obtained for $\mathcal{F}_{\mu \Delta}$:  
\begin{align}
\FmD &=  \sum_k \frac{\Delta z_k \sin^2 k}{\mathcal{E}_k^4}.
\label{eq:QFIM-ele}
\end{align}
Unlike classical Fisher Information, which is additive, quantum Fisher Information is super-additive. {Hence, the quantity of prime interest in the case of multiparameter estimation is the system-size scaling of the ultimate precision estimator $\mathcal{G}$ in Eq.~(7). By calculating the limits of QFIM elements as $\Delta \to 0$ and $\mu \to 2$, we obtain closed form expressions of the leading order scalings as, 
\begin{align}
\FmD &\sim \frac{\Delta }{\pi^4}L^4 \nonumber\\
\mathcal{G} &\sim \frac{(\pi^2 - 8) \Delta^2 }{\pi^8}L^6. 
\label{mu2del0-QFIM}
\end{align}
Thus, we observe super-Heisenberg scaling in all the QFIM elements except $\FDD$ for which we showed that it saturates the Heisenberg limit.}  These scalings of the QFIM elements are remarkable results establishing the importance of the multicritical point.   The multicritical point can be approached via different paths (Fig.~\ref{fig:schematic}) in the two-dimensional parametric plane of $\mu$ and $\Delta$. It turns out that the enhancement is a genuinely multi-parameter effect that is rooted in the coexistence of gap-to-gapless transition and symmetry-breaking.

\begin{figure*}[t]
    \centering
    \includegraphics[scale = 0.3]{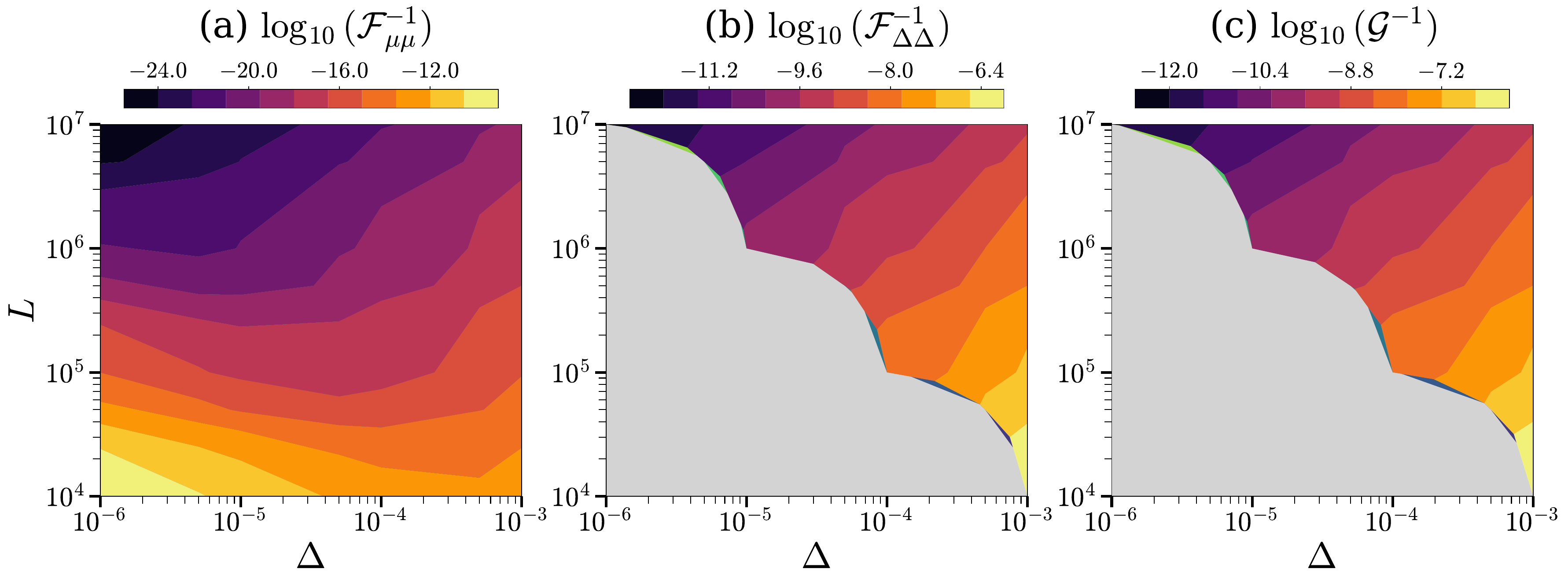}
    \caption{
    \textcolor{black}
    {\textit{\bf{Region of good sensor}}: The QFIM elements \textbf{(a)} $\mathcal{F}_{\mu \mu}^{-1}$, \textbf{(b)} $\mathcal{F}^{-1}_{\Delta \Delta}$ and \textbf{(c)} $\mathcal{G}^{-1}$ are presented. The value of $\mu$ is fixed at $\mu = 2$. According to quantum Cram\'er-Rao bound, $\delta {\theta}^2 \geq (M F_{\theta})^{-1}$, where $M$ is the number of measurements. Thus for a effective sensor, the values of $(M F_{\theta})^{-1}$ should be at-least lesser by an order than the parameter $\theta$ that is being sensed, viz. $1/\sqrt{MF_{\theta}} \ll |\theta|$. Typically, the experiments are repeated for few thousand times, i.e., $M=10^3-10^4$. Hence, practical parametric regime for effective sensing can be identified by validity of the constraint, ${F_{\theta}}^{-1/2} < |\theta|$. We shade the regions gray, where this is not the case. In (a) and (b), we consider  single-parameter estimation case, while in (c) we consider the multi-parameter estimation.} 
}
    \label{fig:QCRB}
\end{figure*}

It is imperative that while changing both the parameters simultaneously in an adiabatic manner, it is done in such a fashion that the parameter $\mu$ does not follow the critical $\Delta=0$, along which $\mathcal{G}$ vanishes. Hence, the variation in $\mu$ must be guided entirely within the symmetry-broken phase with non-vanishing $\Delta$, and such that it encounter a gap-to-gapless transition. However, the limit $\Delta \to 0$ involves symmetry-breaking, as discussed before. The combined effects of the gap-to-gapless transition and the symmetry-breaking gives rise to the  super-Heisenberg scaling in $\mathcal{G}$.

In Fig.~\ref{fig:G}(a), we present the variation of $\mathcal{G}$ with respect to the parameters $\mu$ and $\Delta$ for system-size $L= 1000$. This is a representative case, and we observe similar behavior for other system-sizes as well. We observe that the value of $\mathcal{G}$ is maximum around the critical point of $\mu = 2$. The nature of $\mathcal{G}$ is similar to that of $\FDD$ in the topological regime ($\mu<2$) and is similar to $\Fmu$ in the trivial phase ($\mu>2$). 
We provide the finite-size scaling of $\mathcal{G}$ in Fig.~\ref{fig:G}(b), where we show that $\mathcal{G}$ has super-HL of $\beta = 6$ at $\Delta = 10^{-7}$ but a SQL scaling at a larger $\Delta = 0.7$. In Fig.~\ref{fig:G}(c) we provide the change in the scaling exponent of $\mathcal{G}$ with change in $\Delta$. The range of values of $\Delta$ can be divided into three subparts: small $\Delta$ region where $\mathcal{G}$ has a super-HL scaling of 6. In the intermediate $\Delta$ range the scaling exponent transitions from super-HL to SQL value of $\beta = 1$. In the large $\Delta$ region the scaling exponent saturates to SQL.


\begin{figure}[h]
    \centering
    \includegraphics[scale = 0.38]{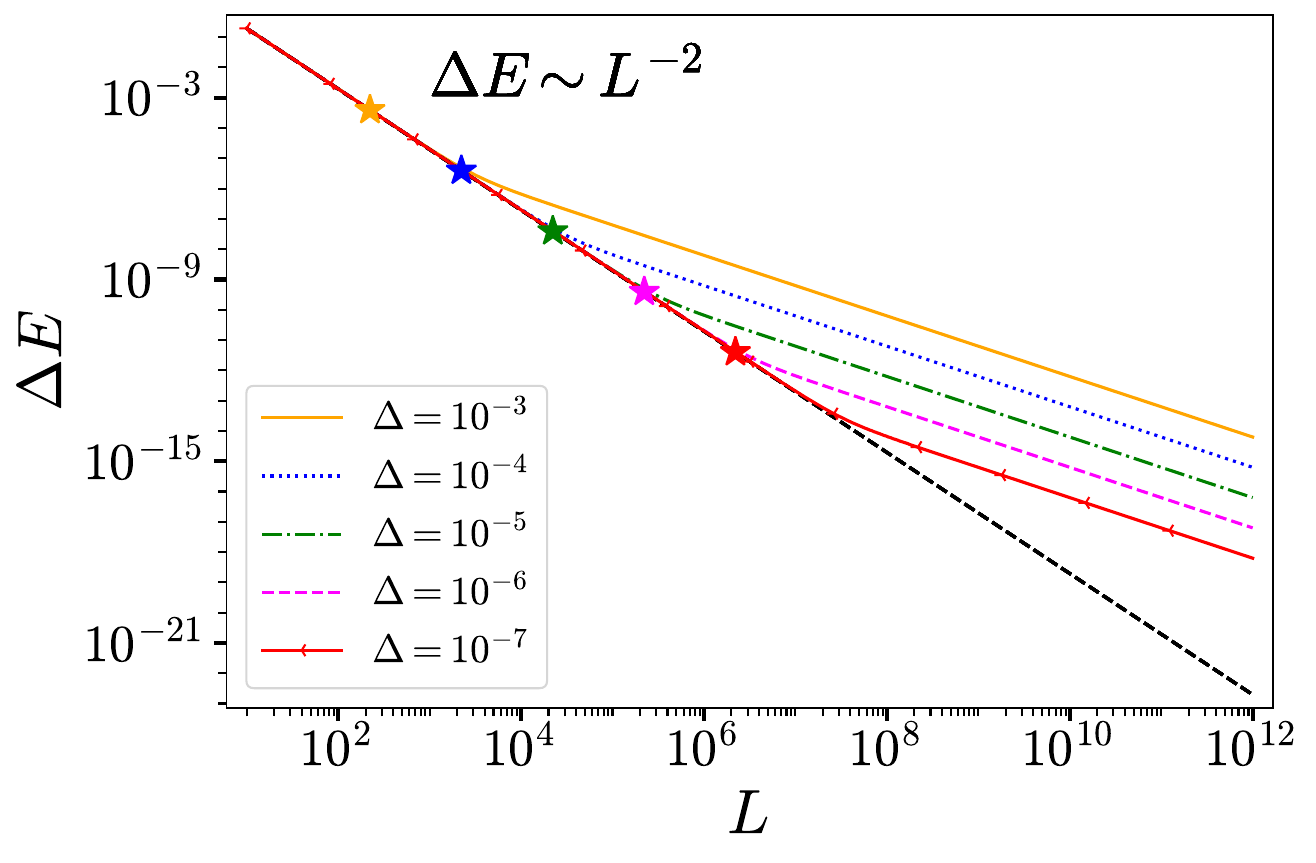}
    \caption{
    {\textit{\bf{Scaling of energy gap $\Delta E$ with system-size}:} The ground state energy gap $\Delta E$ is presented for various values of $\Delta$ at $\mu = 2$. For smaller system sizes, $\Delta E \sim L^{-2}$, but at larger system size, $\Delta E \sim L^{-1}$. A similar change in scaling is also observed in quantum Fisher information where $\mathcal{F}_{\mu\mu}$ transitions from a scaling of $L^6$ to $L^2$ as the system size increases. We define a system-size $L_c$, at which the quantity $\mathcal{F}_{\mu \mu}$ deviates from the $L^6$ scaling by $1\%$. We also observe that $L_c \propto \Delta^{-1}$. We represent the $L_c$ with stars in the figure.  }     
}
    \label{fig:DeltaEvsL}
\end{figure}
\section{Practical Constraints}
\label{Sec4}
{{In this section, we perform critical assessment of the proposed sensors in order to infer the their limitations after taking into account the practical constraints. In particular, we perform following analysis: firstly, we investigate the parameter regime, where relative strength of the signal-to-noise ratio is considerable, for which we need to closely inspect the absolute values of the QFIs with respect to the strength of the unknown parameters;  secondly, given that true adiabaticity is an idealistic scenario and parameter maneuvering must include a finite time ramping in reality, we perform an analysis to see if quantum advantage persists even after state preparation time is accounted for.

\emph{Parameter regime for good sensors.} From the quantum Cramér–Rao bound, the uncertainty in estimating $\theta$ satisfies $\delta \theta^2 \ge (M F_{\theta})^{-1}$, where $M$ is the number of measurement repetitions. For the sensor to perform well, the sensitivity limit should be much smaller than the actual parameter value being measured, implying $1/\sqrt{M F_{\theta}} \ll |\theta|$. In typical experiments, the measurement is repeated about $10^3$--$10^4$ times. Therefore, a practical operational regime for reliable sensing can be identified using the simpler condition, $1/\sqrt{F_{\theta}} < |\theta|$.} 

{The QFIM elements are independent of $\mu$ in the leading order at multicriticality, while it is proportional to $\Delta^2$ in the leading order. The QFIM elements decrease as $\Delta$ decreases and increase as system-size $L$ increases. This leads to a trade-off, and we need to identify the parameter regime for which the system acts as good sensors.}

The parameter $\Delta$ is varied across several orders of magnitude. We investigate the variation of qCRb varies with respect to it.  In Fig.~\ref{fig:QCRB}, we have presented the qCRb-s: (a) $\Fmu^{-1}$ for the parameter estimation of $\mu$, (b) $\FDD^{-1}$ for the  parameter estimation of $\Delta$, and (c) $\mathcal{G}^{-1}$ for the two-parameter estimation of $\mu$ and $\Delta$. We mask the portion gray, where the respective qCRb-s are larger than the corresponding parameter values. { In Fig.~\ref{fig:QCRB}(a), the whole parameter region shown is effective for sensing $\mu$, while in Fig.~\ref{fig:QCRB}(b) and (c) the masked regions do not provide advantage in sensing.}  
The colored region are effective region for sensing. We would like to point that this analysis is for single iteration of the sensing protocol, for multiple runs of the protocol, the precision improves further as it is inversely proportional to the number of times the protocol is carried out.

\emph{Effects of state preparation time}. { We investigate the sensor performance while explicitly accounting for the probe-state preparation time in these criticality-based adiabatic quantum sensors.} Hence, the ratio of QFI and time, viz., $F_Q/t$ is our quantity of interest instead of the QFI $F_Q$. For adiabatic time evolution of the ground state, it is essential that the time $t$ for which the evolution takes place is of the order of the inverse of ground state energy gap $t \sim (\Delta E/\hbar)^{-1}$. 
{Setting $\mu = 2$, we get 
\begin{align*}
\Delta E &= E_1 - E_0 = 2\mathcal{E}_k^\text{min} \nonumber\\ 
&=  4\sqrt{\left(1+\cos k_{\min}\right)^2 + \frac{1}{4} \Delta^2 \sin^2k_{\min}} 
\end{align*}
It turns out that $k_{\min} = \pi - \frac{\pi}{L}$ at $\mu = 2$. 
Thus, expanding $\Delta E$ at $k_{\min}$, we get
\begin{align*}
    \Delta E = 4\sqrt{\frac{\pi^2}{L^2}\left( \frac{\Delta^2}{4} + \frac{\pi^2}{L^2}\right)}.
\end{align*}
As discussed in the previous section, there exists a finite system size $L_c$, beyond which the QFI of the ground state of the state transition from $L^6$ to $L^2$. A similar feature is observed for $\Delta E$ as well, shown in Fig.~\ref{fig:DeltaEvsL}. When $\Delta \gg 2\pi/L$, $\Delta E \sim L^{-1}$ whereas when $\Delta \ll 2\pi/L$, we have $\Delta E \sim L^{-2}$. 
}
{
We find in the limit $\Delta \ll L^{-1}$, $\Delta E \sim L^{-2}$. Thus, we have $\Fmu/t \sim L^4$ (super-HL), $\FmD/t \sim L^2$ (HL) and $\mathcal{G}/t \sim L^4$ (super-HL). Although, $\FDD/t \sim L^0$ loses any system-size scaling. 
On the other hand, when  $\Delta \gg L^{-1}$, we have $\Delta E \sim L^{-1}$, and $\mathcal{F}_{\mu\mu} \sim L^2 \Rightarrow \mathcal{F}_{\mu\mu}/t \sim L$, hence, we achieve the standard quantum limit. Here, we are far from the critical region of parameter $\Delta$, hence there is no enhancement in QFI of $\Delta$ in this region. We would like to comment, as it's obvious, that better system-size scaling of the QFI can be obtained by increasing preparation time $t$ in our protocol for finite-system sizes.}

\vspace{-2mm}
\section{Discussions}
\label{Conc}
{This work demonstrates both fundamental and  practical contribution to quantum metrology. Fundamentally, we identify symmetry-breaking can be an independent resource resulting in Heisenberg-limited scaling without invoking gap closing in QMB critical sensing. Practically, we show that quantum enhancements persists even after accounting for practical constraints, such as realistic state-preparation times.

In context of single-parameter estimation, we showed that preparing near the multicritical point of the Kitaev chain and varying $\Delta$ along the gapless line $\mu=2$, results in Heisenberg-limited scaling of $\FDD$ without invoking a gap-to-gapless transition. This demonstrated that symmetry breaking alone can also serve as a metrological resource. In the multiparameter setting, super-Heisenberg scaling of the combined estimator $\mathcal{G}$ arises from the coexistence of symmetry breaking and gap closing near the multicritical point. Importantly, from a practical standpoint, we suggested viable operating windows in the parametric regime, where the proposed sensors operates most efficiently. Moreover, we showed that a definitive quantum advantage and super-extensive scalings persists when the adiabatic state-preparation costs are accounted for.}

We would like to point out that, the Kitaev model considered here has a correspondence with the widely studied XY model on a one-dimensional spin lattice \cite{Mbeng2024} and the topological critical points have corresponding critical points in the XY model which are physically critical points of a second-order phase transition. Hence, our work addresses a more generic setting beyond the system considered.

Although in our analysis, we do consider the effect of adiabatic preparation of the probe state Ref.~\cite{Chaves2013,Brask2015,Smirne2016,Albarelli2018,He2023, reviewMontenegro2024}, it would be interesting to consider the effect of various other ways to prepare the probe state as discussed in Refs.~\cite{PhysRevLett.132.100803, Gietka2021}. It will be interesting to pursue further investigations in future, such as understanding the effects of long-range tunneling or long-range pairing on the sensing protocols proposed in this work.

\section*{Appendix}
\appendix

\section{Derivation of $\mathcal{F}_{ab}$}
\label{app:A}
The elements of QFIM for a pure state are given by,
\begin{align}
    \mathcal{F}_{ab} = 4\text{Re}(\langle \partial_a \psi|\partial_b \psi \rangle - \langle \partial_a \psi|\psi \rangle\langle \psi|\partial_b \psi \rangle).
\end{align}
For the Kitaev model the ground state is of the form:
\begin{align}
    |\Psi_G\rangle = \prod_{k = 1}^{L/2} \gamma_k^\dagger |0\rangle,  
    \label{eq:probe-state}
\end{align}
where $k$ stands for the momentum states, $k = (2i-1)\pi/L$, for $ i = 1,..,L/2$. Here we do the calculations considering the anti-periodic boundary conditions (ABC), but these results holds for periodic boundary conditions (PBC) as well. For brevity, we mark $k$ via $i$ in the product above and 
$\gamma_k^\dagger = u_k + v_k c_k^\dagger c_{-k}^\dagger .$ Furthermore, 
\begin{align}
\label{uk_vk_def}
    u_k &= \frac{\Ek + z_k}{\sqrt{2\Ek(\Ek + z_k)}}\nonumber, \\
    v_k &= \frac{iy_k}{\sqrt{2\Ek(\Ek + z_k)}}.
\end{align}
Here $\Ek = 2\sqrt{(\cos k + \frac{\mu}{2})^2 + \frac{1}{4}\Delta^2 \sin^2k}$, $y_k = -\Delta \sin k$ and $z_k = -\mu - 2\cos k$. Using the form,
\begin{align}
    |\partial_a \Psi_G\rangle = \sum_{k = 1}^{L/2} \gamma^\dagger_1 ... (\partial_a u_k + \partial_a v_k c_k^\dagger c_{-k}^\dagger)...\gamma_{L/2}^\dagger|0\rangle.
\end{align}
\begin{align}
    \Rightarrow \langle \partial_a \Psi_G |\partial_b \Psi_G\rangle &= \sum_{k = 1}^{L/2} \partial_a u_k^* \partial_b u_k + \partial_a v_k^* \partial_b v_k . \nonumber \\
   \Rightarrow \langle \partial_a \Psi_G | \Psi_G\rangle &= \sum_{k = 1}^{L/2}u_k\partial_a u_k^*  + v_k\partial_a v_k^* .
\end{align}
Using this form, we arrive at the final expression:
\begin{align}
    &\mathcal{F}_{ab} = 4\sum_{k = 1}^{L/2}\text{Re}\left[\partial_a u_k^* \partial_b u_k + \partial_a v_k^* \partial_b v_k \right] \nonumber \\ &+ 4\text{Re}\left[\left(\sum_{k = 1}^{L/2}u_k\partial_a u_k^*  + v_k\partial_a v_k^*  \right)\left(\sum_{k = 1}^{L/2} u_k^* \partial_b u_k  + v_k^* \partial_b v_k  \right)\right].
\end{align}
We note that $\sum_{k = 1}^{L/2}u_k\partial_a u_k^*  + v_k\partial_a v_k^*   = 0$ using $|u_k|^2 + |v_k|^2 = 1$ and Eq.~ \eqref{uk_vk_def}. Thus, we get:
\begin{align}
    \mathcal{F}_{ab} &= 4\sum_{k = 1}^{L/2}\text{Re}\left[\partial_a u_k^* \partial_b u_k + \partial_a v_k^* \partial_b v_k \right].
\end{align}

We find that:
\begin{align}
    \partial_\mu u_k &= \frac{(-\Ek + z_k)\sqrt{\Ek(\Ek+z_k)}}{2\sqrt{2} \Ek^3},\nonumber \\
    \partial_\mu v_k &= \frac{iy_k\sqrt{\Ek(\Ek+z_k)}}{2\sqrt{2}\Ek^3},\nonumber \\
    \partial_\Delta u_k &= \frac{-y_k^2z_k}{2\sqrt{2}\Delta \Ek^{5/2} \sqrt{\Ek+z_k}},\nonumber \\
    \partial_\Delta v_k &= \frac{iy_kz_k(\Ek+z_k)}{2\sqrt{2}\Delta \Ek^{5/2} \sqrt{\Ek+z_k}}.
\label{derivatives}
\end{align}
Subsequenty, we get,
\begin{align}
    \Fmu &= \sum_k \frac{\Delta^2 \sin^2 k}{\Ek^4},\nonumber \\
    \FmD &=  \sum_k \frac{\Delta z_k \sin^2 k}{\Ek^4},\nonumber \\
    \FDD &= \sum_k \frac{z_k^2 \sin^2k }{\Ek^4},
    \label{eq:FI-analytic}
\end{align}
where $k = (2i-1)\pi/L$ and sum is over the values $i = {1,2,..,L/2}$. We corroborate the expressions obtained in Eq.~\eqref{eq:FI-analytic} by comparing them with  numerically obtained values of the elements of the QFI in Fig.~\ref{fig:fig7}.

\begin{figure}
    \centering
    \includegraphics[scale = 0.34]{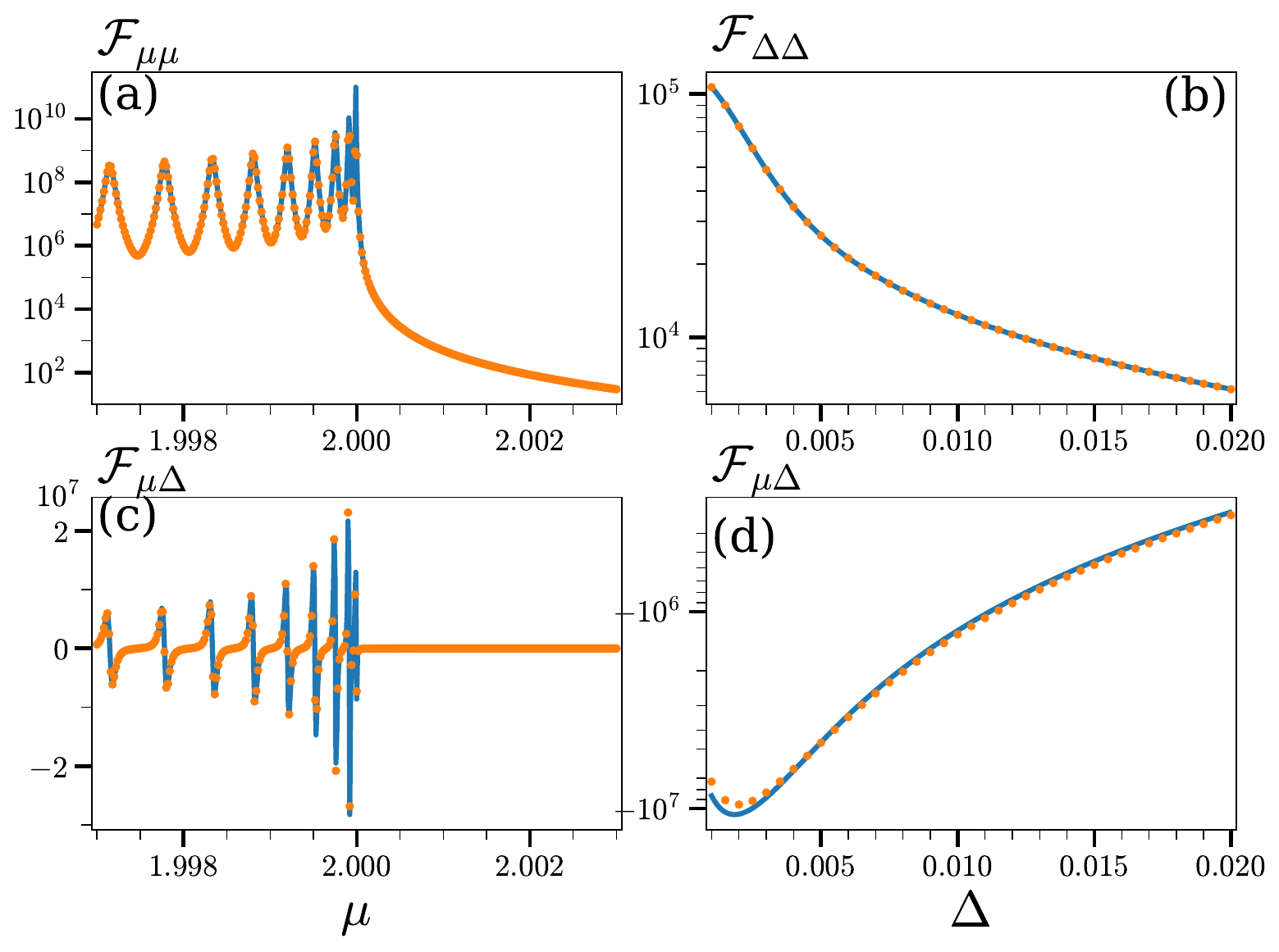}
    \caption{\textbf{\textit{Comparison between analytic expression as written in Eq.~\eqref{eq:FI-analytic} of the QFIM and  their corresponding values obtained numerically}}: The elements of QFIM are presented with change in $\mu$ or $\Delta$, i.e., (a) $\Fmu$ against $\mu$, (b) $\FDD$ against $\Delta$, (c)$\FmD$ against $\mu$ and (d) $\FmD$ against $\Delta$. This is a representative case of system-size $L= 1000$, where $\Delta = 0.001$ in (a) and (c), while $\mu = 2.0$ in (b) and (d). The blue lines are obtained from the Eq.~\eqref{eq:FI-analytic}, while the orange circular markers are obtained numerically for the Kitaev chain with anti-periodic boundary condition. The y-axes of (a),(b) and (d) are in log-scale while the y-axis of (c) is in linear-scale. The x-axes of all four are in linear-scale.}
    \label{fig:fig7}
\end{figure}
\begin{figure}[!t]
    \centering
    \includegraphics[scale = 0.34]{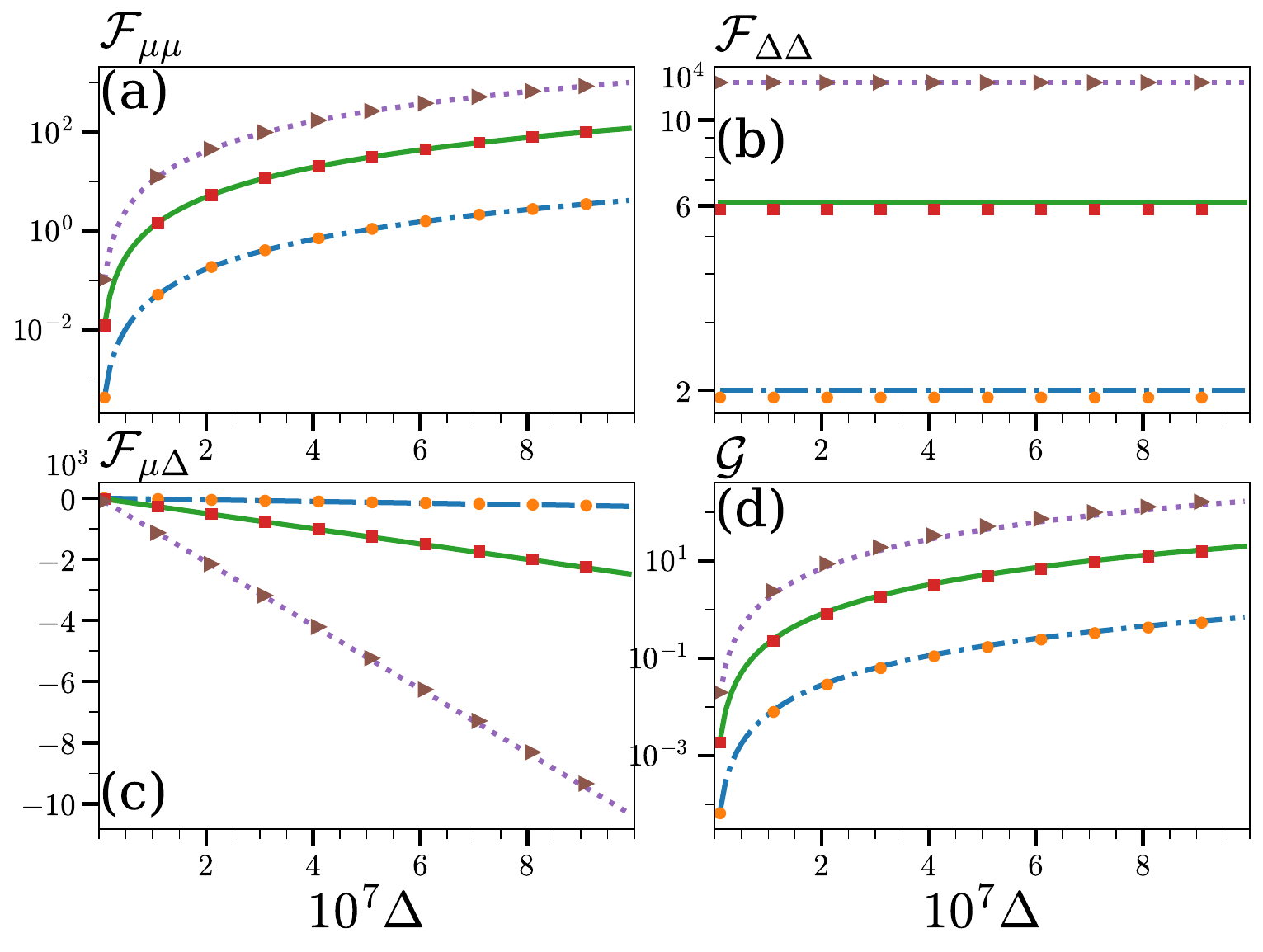}
    \caption{\textbf{\textit{Comparison between Eq.~\eqref{eq:FI-analytic} and  Eq.~\eqref{eq:FI-approx}}}: The analytic form of the elements of the Quantum Fisher Information matrix: (a) $\Fmu$, (b) $\FDD$, (c) $\FmD$ and (d) $\mathcal{G}$ are presented with respect to $\Delta$ which is in the range $\Delta \in [10^{-8},10^{-6}]$, keeping a fixed $\mu = 2$. We numerically confirm that these quantities (from Eq.~\eqref{eq:FI-analytic}) and its approximations (from Eq.~\eqref{eq:FI-approx}) in the small $\Delta$ regime matches. The markers indicate the approximated functions 
 as written in Eq.~\eqref{eq:FI-approx} and the lines correspond to Eq.~\eqref{eq:FI-analytic}. These are plotted for system sizes $L = 400$ (circular markers, dot-dashed line) , $700$ (square markers, solid line) and $1000$ (triangular markers, dotted line). The markers and line-styles are uniform across all the four sub-figures. The y-axes are in log-scale in (a),(b) and (d) and in linear-scale in (c). The x-axis is in linear-scale and the labels are multiplied by $10^7$ for convenience.} 
    \label{fig:fig8}
\end{figure}

Let's find out the scaling exponents, when we set both $\Delta \rightarrow 0$ and $\mu \rightarrow 2$. We have
\begin{align}
    \Fmu &= \sum_k \frac{\Delta^2 \sin^2k}{(4(\cos k + \frac{\mu}{2})^2 + \Delta^2 \sin^2k)^2} \nonumber, \\
    &\approx \sum_k \frac{\Delta^2 \sin^2k}{16(\cos k + 1)^4 } + \mathcal{O}(\Delta^4).
\end{align}
Now, as we can see the numerator tends to zero. Therefore, largest contribution of this sum comes from the terms where the denominator also tends to zero, i.e., $k\rightarrow \pi$. This is achieved by the last term in the sum, which is given by setting $i = L/2$ and therefore $k = \pi - \pi/L$. We thus get, 
\begin{align}
    \Fmu  \approx  \frac{\Delta^2 \sin^2(\pi - \frac{\pi}{L})}{16(\cos (\pi - \frac{\pi}{L}) + 1)^4 } - \frac{\Delta^4 \sin^4(\pi - \frac{\pi}{L})}{64(\cos (\pi - \frac{\pi}{L}) + 1)^6}.
    \label{eq:fmu_approx}
\end{align}
In the limit of large $L$, we get
\begin{align}
    \Fmu  \approx  \frac{\Delta^2 L^6}{\pi^6 } + \mathcal{O}(\Delta^4).
\end{align}
Thus, we have $\Fmu \sim L^6$ for $\Delta \rightarrow 0$ and $\mu \rightarrow 2$. Using similar arguments it can be easily shown that $\Fmu \sim L^2$, when $\mu \rightarrow 2$ and $|\Delta| \gg 0$.

\begin{figure}[t]
    \centering
    \includegraphics[scale= 0.4]{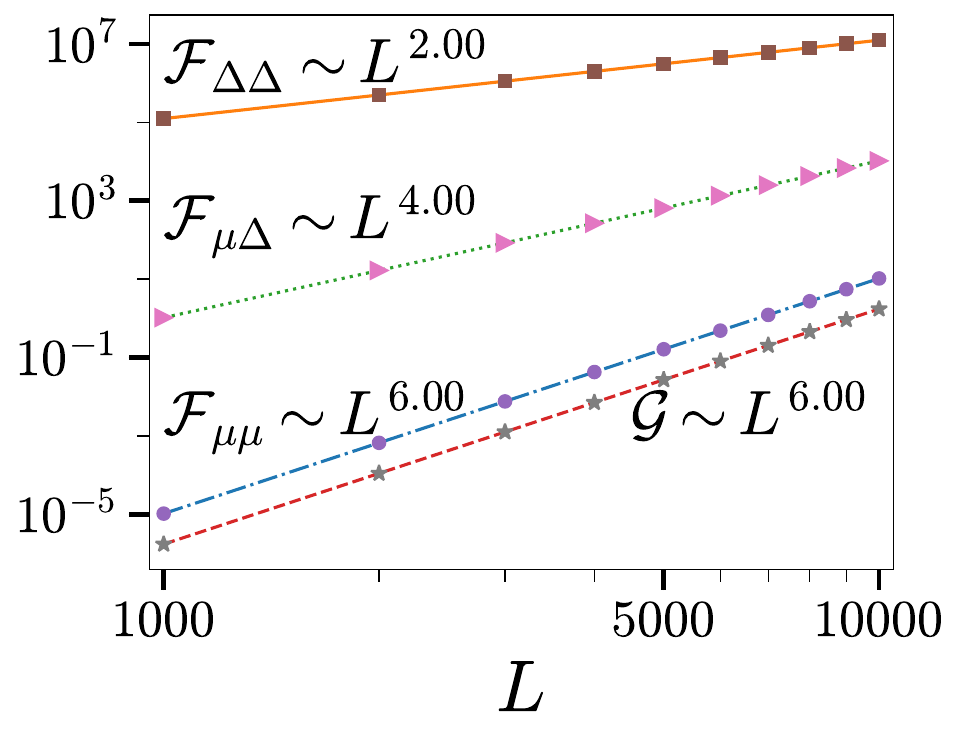}
    \caption{\textbf{\textit{Finite-size scaling of elements QFIM and $\mathcal{G}$}}: We evaluate $\Fmu$ (circular markers), $\FDD$ (square markers), $-\FmD$ (triangular markers) and $\mathcal{G}$ (star markers) at the point $\Delta = 10^{-10}$ and $\mu = 2.0$ using Eq.~\eqref{eq:FI-analytic} for system-sizes from $L = 1000$ to $10000$. The scaling exponents match with the ones obtained in Eq.~\eqref{eq:FI-approx}. The lines represent the best fits.  }
    \label{fig:scaling-analytical}
\end{figure}

\begin{figure}[t]
    \centering
    \includegraphics[scale = 0.2]{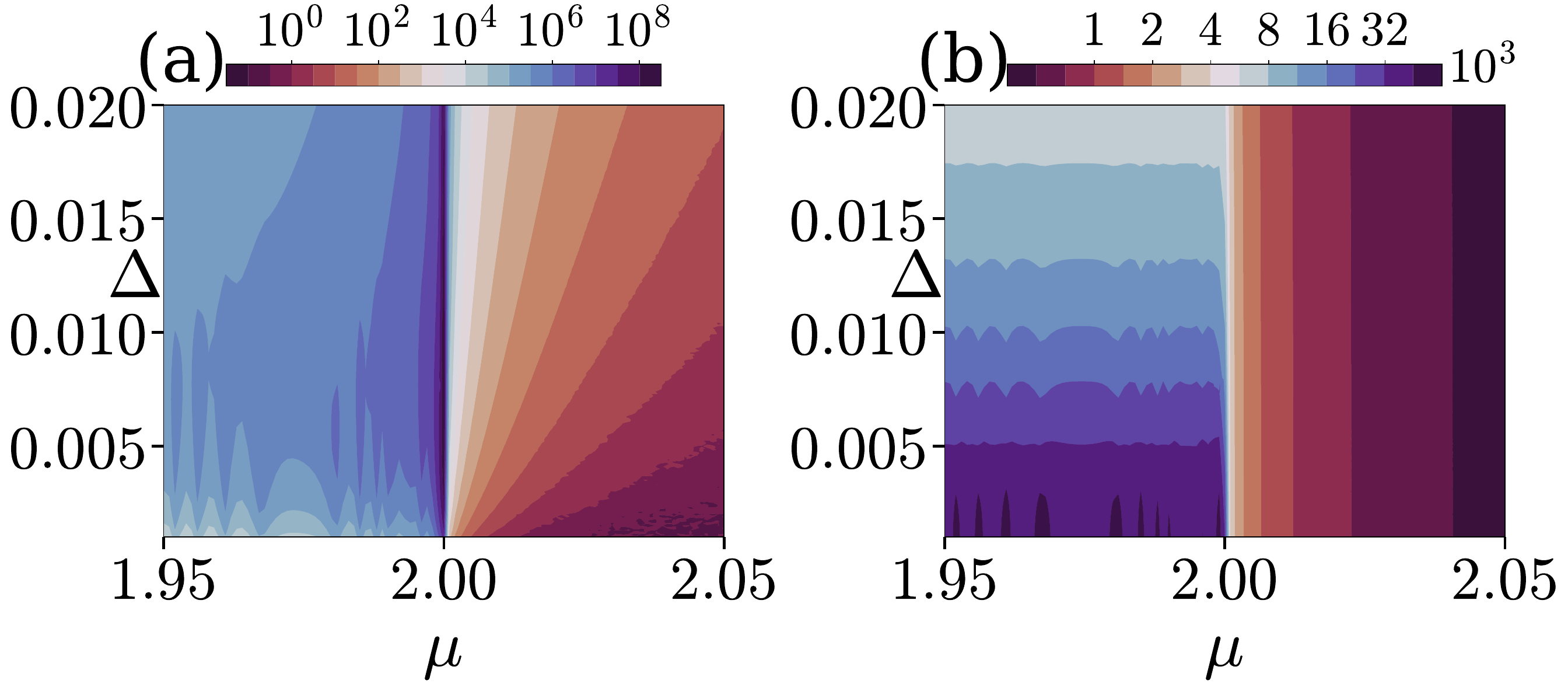}
    \caption{\textit{\bf{Elements of the Quantum Fisher Information matrix}}: The QFIM elements \textbf{(a)} $\mathcal{F}_{\mu \mu}$ and \textbf{(b)} $\mathcal{F}_{\Delta \Delta}$ are presented with respect to the tuning parameters $\mu$ and $\Delta$ in the vicinity of the critical points of either axes. The probe considered is the ground state of the Kitaev model on a one-dimensional lattice with system size of $L = 1000$. Note that in (b) the Fisher information is in the range of $10^{3}$. In both figures, the elements of QFIM are presented in log-scale.
}
    \label{fig:2dN1Ka}
\end{figure}

\begin{figure}[!h]
    \centering
    \includegraphics[scale = 0.38]{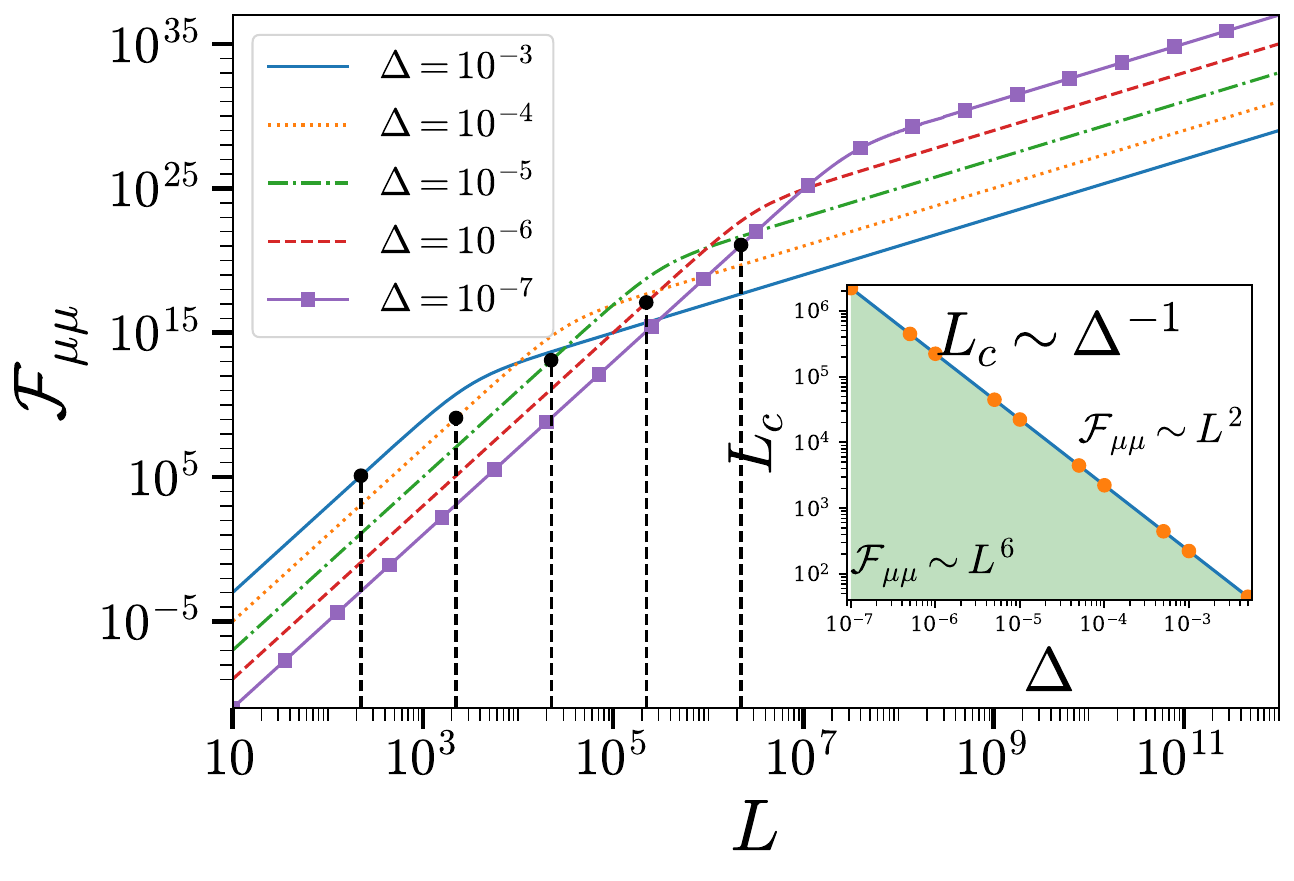}
    \caption{\textcolor{black}{\textit{\bf{Validity of the small $\Delta$ expansion}:} We observe that the system-size scaling of $\mathcal{F}_{\mu \mu}$ undergoes a gradual transition from $L^6$ to $L^2$ as system-size $L$ increases. We define a specific system-size $L_c$ (represented by black dots and dotted-vertical lines), at which value the actual $\mathcal{F}_{\mu \mu}$ deviates from the corresponding $L^6$ fit by $1\%$. We realize that $L_c$ increases as $\Delta$ decreases. This validates our assumption of considering $\Delta \rightarrow 0$ limit as we can always find a finite $L_c$ for a corresponding non-zero $\Delta$. In the inset, we fit the $L_c$ with $\Delta$ and find the fit : $L_c \sim \Delta^{-1}$. }     
}
    \label{fig:L-change}
\end{figure}


\emph{Validity of $\Fmu \sim L^6$} : 
In Fig.~\ref{fig:L-change}, we observe that for a specific value of $\Delta$, there exist a $L$ value, up to which we observe $\Fmu \sim L^6$ and beyond which $\Fmu \sim L^2$. In the Fig.~\ref{fig:L-change}, we present this for different values of $\Delta$. We define a specific system-size $L_c$, which is the system-size at which the actual value of $\Fmu$ deviates by $1\%$ from the extrapolated value it would have had with $\Fmu \sim L^6$ fit. 
The black dotted lines on the $L$-axis, 
corresponds to the different critical lengths $L_c$-s obtained for various values of $\Delta$-s. 
In the inset we observe that they fit as $L_c \sim \Delta^{-1}$. The green shaded region corresponds to the region where $\Fmu \sim L^6$. 
Hence, as $\Delta \rightarrow 0$, we have $L_c \rightarrow \infty$. This corroborates our claim that as $\Delta \rightarrow 0$, we have $\Fmu \sim L^6$.

For $\FDD$ in the limit $\Delta \rightarrow 0$ and $\mu \rightarrow 2$, we have
\begin{align}
    \FDD  \approx \sum_k \frac{1}{4} \tan^2\left(\frac{k}{2}\right) + \mathcal{O}(\Delta^2) .
\end{align}
We observe that the larger values of $k$ dominate this sum. By considering the contributions to the sum in descending order of the  $k = \pi - \frac{\pi}{L}$, one finds
\begin{align}
    \FDD  &\approx  \frac{1}{4} \tan^2\left(\frac{\pi}{2} - \frac{\pi}{2L}\right) + \frac{1}{4} \tan^2\left(\frac{\pi}{2} - \frac{3\pi}{2L}\right) + .... \nonumber\\
     &=  \frac{1}{4} \cot^2\left(\frac{\pi}{2L}\right) + \frac{1}{4} \cot^2\left( \frac{3\pi}{2L}\right) + .... \nonumber\\
     &\approx  \frac{L^2}{\pi^2} \left(1 + \frac{1}{9} + \frac{1}{25}.... + \frac{1}{m^2}\right).
     \label{eq:FDD-zeta}
\end{align}
Here, we terminate the series at some $m\ll L/2$. Now we note that the Riemann zeta function at $n = 2$ is given by 
\begin{align}
    &\zeta(2) = \sum_{n = 1}^\infty \frac{1}{n^2} 
    =\frac{\pi^2}{6}, \nonumber \\
    \Rightarrow &\left(1  + \frac{1}{9} + \frac{1}{25} +.... \right) = \frac{3}{4}\zeta(2) = \frac{\pi^2}{8}.
    \label{eq:zeta-odd}
\end{align}
For very large $L$ and consequently large $m$, we can approximate the sum in bracket of Eq.~\eqref{eq:FDD-zeta} with Eq.~\eqref{eq:zeta-odd}. We thus get,
\begin{align}
    \FDD \approx \frac{L^2}{8} + \mathcal{O}(\Delta^2).
\end{align}
Hence, we have $\FDD \sim L^2$ for $\Delta \rightarrow 0$ and $\mu \rightarrow 2$. Following a similar line of reasoning as Eq.~\eqref{eq:fmu_approx}, we get
\begin{align}
    \FmD \approx - \frac{\Delta L^4}{\pi^4}  + \mathcal{O}(\Delta^3).
\end{align}

Thus, we have $-\FmD \sim L^4$ for $\Delta \rightarrow 0$ and $\mu \rightarrow 2$.
Now, for the case of $\mathcal{G}$, it can be shown that 
\begin{align}
    \mathcal{G} \approx \frac{(\pi^2 - 8)\Delta^2}{\pi^8}L^6 +  \mathcal{O}(\Delta^4).
\end{align}
Thus, we have $\mathcal{G} \sim L^6$ for $\Delta \rightarrow 0$ and $\mu \rightarrow 2$. In summary, one obtains,
\begin{align}
     \Fmu  &\approx  \frac{\Delta^2 }{\pi^6 }L^6, \quad 
    \FDD \approx \frac{1}{8}L^2, \nonumber \\
    \FmD &\approx - \frac{\Delta }{\pi^4}L^4, \quad
    \mathcal{G} \approx \frac{(\pi^2 - 8)\Delta^2}{\pi^8}L^6.
    \label{eq:FI-approx}
\end{align}

In Fig.~\ref{fig:fig8}, we check the validity of Eq.~\eqref{eq:FI-approx} by comparing it with Eq.~\eqref{eq:FI-analytic} in the regime of  $\Delta \to 0$ and $\mu = 2$. Moreover, in Fig.~\ref{fig:scaling-analytical} we present the finite-size scaling using Eq.~\eqref{eq:FI-analytic} and find that the scaling exponents match with the ones analytically obtained in Eq.~\eqref{eq:FI-approx}.

In Fig.~\ref{fig:2dN1Ka} we present (a) $\Fmu$ and $(b) \FDD$ for system-size $L = 1000$. We observe a similar nature for other system-sizes as well. Fig.~\ref{fig:2dN1Ka} quite clearly shows the transition from topological to trivial phase at $\mu = 2.0$. We see in Fig.~\ref{fig:2dN1Ka}(a) that the system achieves its maximum $\Fmu$ at the point $\mu = 2.0$. The quantity $\Fmu$ gradually decreases as we move further into the trivial phase. 
Due to the apparent fluctuations of $\Fmu$ in the topological phase of the finite-sized systems, particularly in smaller systems, numerical calculations were not pushed in the smaller $\Delta$ regime. We, however, performed the scaling analysis in the trivial phase away from the critical line. In Fig.~\ref{fig:2dN1Ka}(b), we present the component $\FDD$ of QFIM. The figure clearly shows that this quantity increases as we move nearer to $\Delta = 0$ line, across which the system the topological invariant undergoes a discrete jump, i.e., a flip in sign of the winding number, given  $|\mu| \leq 2$.


%

\begin{figure}[t]
    \centering
    \includegraphics[scale = 0.37]{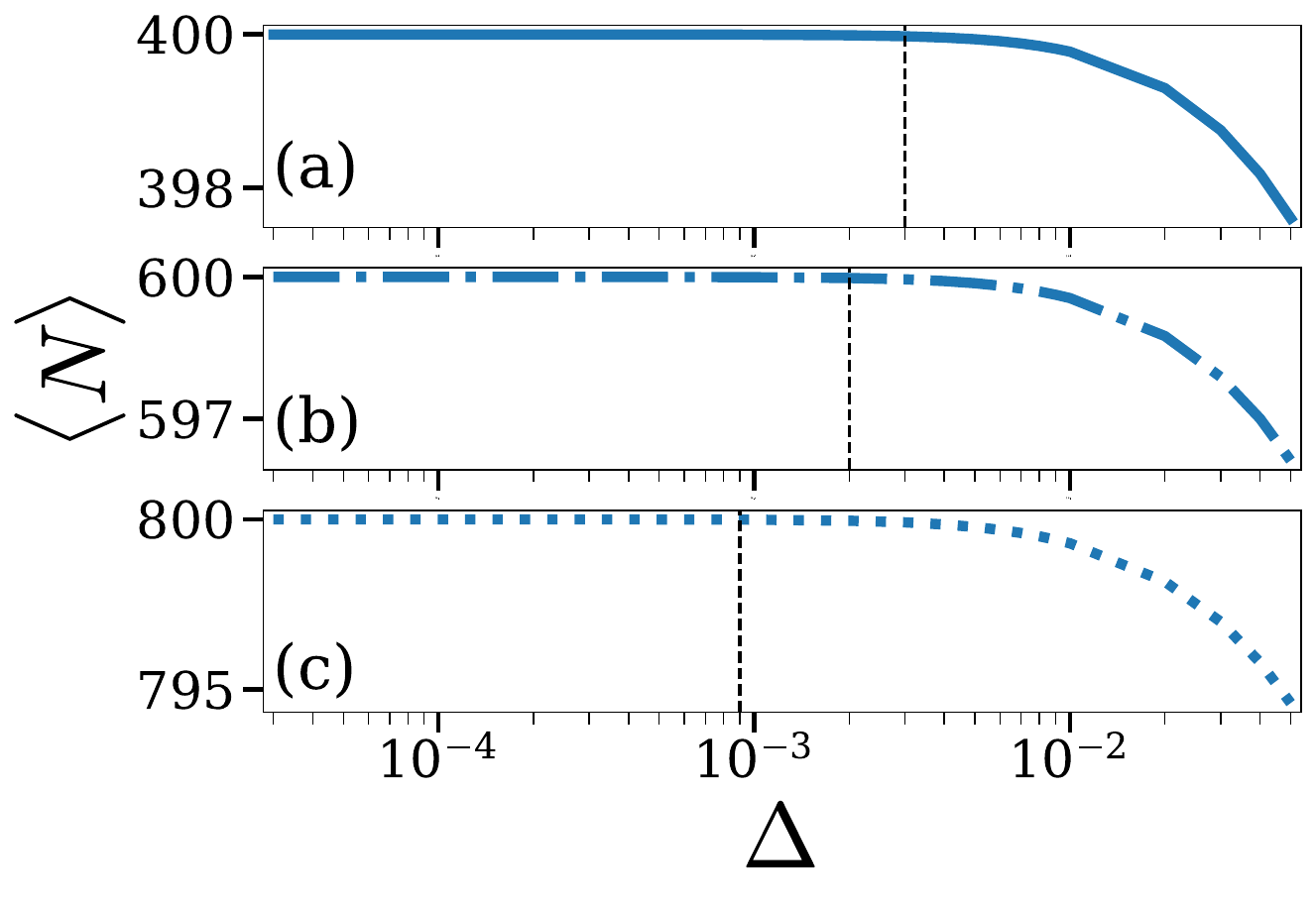}
    \caption{\textit{\bf{Change in the average particle number $\langle N\rangle$ with change in $\Delta$}}: The average particle number of the probe state Eq.~\eqref{eq:probe-state} is numerically obtained and is presented with a fixed $\mu = 2$ throughout. This a representative case for three system sizes : (a) $L$ = 400 (dot-dashed line), (b) $L$ = 600 (solid line) and (c) $L$ = 800 (dotted line). 
}
    \label{fig:avg_N}
\end{figure}
\section{Average particle number in the ground state}
\label{avg-num-state}
{In Fig.~\ref{fig:avg_N}, we present the average particle number $\langle N \rangle$ of the probe state, i.e., the ground state of the Kitaev model in one dimension for $\mu = 2$. As seen from the Hamiltonian of the Kitaev model, the $p$-wave superconductivity term with coupling $\Delta$ makes the system particle-number non-conserving. This means that when $\Delta$ is set to a non-zero value, the ground state wave-function becomes a superposition of various particle-number sector. When $\Delta$ is set to zero, we note two things: the states with fixed momentum become eigenstates of the Hamiltonian and the value of $\mu$ determines the fixed particle-number sector to which the ground state belongs to. We note that at $\mu = 0$ limit, half of the single-particle states have negative eigenenergies. The many-body ground state of the model is constructed by filling the negative energy states via fermions. Since there is a negative sign with $\mu$ in the Hamiltonian,  more single-particle eigenstates with negative energies becomes available with increasing $\mu$. Thus, occupation number of the ground state increases gradually with increasing $\mu$. On introducing a non-zero $\Delta$, the ground state no longer has a fixed number of particles, rather it is a superposition of many fixed particle-number states. We observe that at very small values of $\Delta$, the ground state has the fully-filled state with almost unit probability because of a small $\Delta$ and a very large $\mu$. As the value of $\Delta$ increases, it becomes energetically more costly to fill up particles than that for smaller $\Delta$. Due to this, the average particle-number decreases with increase of $\Delta$. Hence, physically it turns out that at the gapless multicritical point ($\mu=2$ and $\Delta=0$), the system is in a trivial band-insulating phase, as all available energy states are filled-up by the fermions. A sufficiently small value of the symmetry-breaking term $\Delta$, i.e., $\Delta^*$, ensures the system's  transition to a critical $p$-wave superconducting phase. This, in turn, facilitates the quantum-enhanced sensing despite the fact that adiabatic variation of the parameter, $\Delta$, follows the gapless line throughout. }

\bibliography{References}

\end{document}